\documentclass[nofootinbib]{revtex4}

\usepackage{graphicx}

\begin{document}
\title{A formally exact field theory for classical systems at equilibrium.}

\author{D. di Caprio \footnote{E-mail: dicaprio@ccr.jussieu.fr} and J.P. Badiali}
\address{Laboratoire d'\'{E}lectrochimie et de Chimie Analytique, UMR 7575,\\
UPMC Univ Paris 06, CNRS, ENSCP,\\
Universit\'e Paris 6, B\^at. F74, B.P. 39\\
4, Place Jussieu, \ 75230 Paris Cedex 05, France}

\begin{abstract}
We propose a formally exact statistical field theory for describing classical
fluids with ingredients similar to those introduced in quantum field theory. We
consider the following essential and related problems :
$i$) how to find the correct field functional (Hamiltonian) which determines the 
partition function, $ii$) how to introduce in a field theory the equivalent of 
the indiscernibility of particles, $iii$) how to test the validity of this 
approach.
We can use a simple Hamiltonian in which a local functional transposes, in terms 
of fields, the equivalent of the indiscernibility of particles.
The diagrammatic expansion and the renormalization of this term is presented.
This corresponds to a non standard problem in Feynman expansion and requires a 
careful investigation.
Then a non-local term associated with an interaction pair potential is 
introduced in the Hamiltonian.
It has been shown that there exists a mapping between this approach and the 
standard statistical mechanics given in terms of Mayer function expansion.
We show on three properties (the chemical potential, the so-called contact 
theorem and the interfacial properties) that in the field theory the
correlations are shifted on non usual quantities. Some perspectives of the 
theory are given.

\end{abstract}


\maketitle

\section{Introduction}
In various domains of physics a description in terms of fields is frequently
used. Hydrodynamics represents a first example in which some fields (densities,
velocities,~...) are introduced for describing the properties of a coarse
grained entity -- the so-called fluid particle.
Later, field theory (FT) has been used as a simple and intuitive tool to predict
the behaviour of complex systems in the domain of soft matter physics
\cite{Safran1,Pincus1,DesCloiseaux,Gompper,deGennes}. These FT
are essentially phenomenological and rely on Hamiltonian functionals introduced
in an ad~hoc manner.
They focus on a mesoscopic scale description and they are based on a more or
less explicit coarse graining procedure. In this context, Hamiltonians are
introduced to describe large classes of phenomena having similar properties
though different in their microscopic details. This further suggests another
type of problems where a FT is extensively used i.e.\ the description of
critical phenomena. Here also the FT is based on the assumption that a detailed
microscopic knowledge of the system is not relevant to describe
its universality class \cite{Kadanoff,ZinnJustin,Amit,Ma}.
And well suited approximations to describe systems with long range
correlations or interactions are introduced.  In relation, field theory
is also used to describe systems with the long ranged Coulomb interactions.
In this case, FT is constructed using the Hubbard-Stratonovich transform of the
standard partition function \cite{KSSHE1,KSSHE3,Parisi}.
Better known as the sine-Gordon transform in the case of the
Coulomb potential, it has given rise to considerable literature 
\cite{Nabutovskii2,
Kholodenko3,Netz2,Netz3,Coalson1,Caillol2,
Brilliantov,Brydges,Yukh1}.
These approaches give an exact description of the systems properties
on a microscopic level. In this respect, they are distinct from the soft matter like
descriptions based on a coarse graining procedure.  The sine-Gordon approaches
introduce an auxiliary field and intricate couplings between fluctuating fields.
In our opinion, this auxiliary field is essentially a mathematical tool,
difficult to associate with any physical quantity. As a result finding meaningful
approximations is rather counterintuitive and the application of such approaches
requires that one be rather cautious \cite{Fisher}.\\
In contrast to these approaches, our main goal is to show that it is possible
to write a FT directly in terms of fields using methods similar to those used in
quantum field theory (QFT). Namely, we show that it is possible to build the
theory around a field, which is a real quantity having a simple physical meaning. 
Moreover we will show that  our FT construction is not only simple and intuitive
but also leads to a complete description at microscopic level. In this paper, we
consider systems at equilibrium.\\
The paper is organized as follows.
In the next Section we present the main requirements which a FT must verify. 
In Section \ref{sec:definingH}, we give the Hamiltonian on which the FT is based :
it contains two terms of different nature. This leads to investigate the Feynman
expansion of a purely local Hamiltonian with an infinite number of coupling
constants. This is developed in Section \ref{ssec:int_representation} where some
important specific aspects of the expansion are shown.  In Section
\ref{sec:InteractingGas} we calculate the partition function in the presence of
an interaction pair potential. In the next Section we establish an exact mapping
between our FT and the standard statistical mechanics given in terms of Mayer
expansion \cite{Balescu}. In Section \ref{sec:discussion} we illustrate on several examples how
our approach may lead to new aspects in statistical physics. Finally, in Section
\ref{sec:conclusion} we give some conclusions and perspectives.

\section{Requirements for a field theory}

Our main assumption is that the partition function $\Xi[\phi]$ of a classical
system can be described exactly via a functional integral defined according
to
\begin{eqnarray}\label{eq:PartFct0}
  \Xi[\phi]= \int\mathcal{D}\phi\exp\left\{-\beta H[\phi]\right\}
\end{eqnarray}
in which $\phi$ is a field, $H[\phi]$ a functional of this field 
which we call Hamiltonian and $\beta=1/(k_B T)$ is the inverse of 
the temperature.\\
To use (\ref{eq:PartFct0}) we must solve several problems. We have to define
$\phi$ and also find an explicit form for $H[\phi]$. It is intuitive
to choose for $\phi$ a real quantity as the density of matter $\rho$ for
instance. This choice represents a fundamental difference from the
Hubbard-Stratonovitch type approaches in which two fields are used, one
being a complex quantity. 
In comparison with the standard description of the liquid state, where the
configuration space spans all possible distributions of the particles, here
$\rho$ is a function defined everywhere in space. As a consequence the number of
degrees of freedom in~(\ref{eq:PartFct0}) is related to the space discretization
required to calculate the functional integral as opposed to the number of
particles.  Then we have to solve a new problem, of how to transpose in FT a
property mimicking the indiscernibility of particles. To relate the FT and the
usual physics we assume that the average of the field, $<\rho>$, corresponds to
the actual density of particles noted $\tilde{\rho}$ \footnote{Quantities
associated with the thermodynamics as opposed to those calculated from the FT
will generally be indicated with a tilde.}.

In addition to the indiscernibility of particles, the so-called classical
statistical mechanics contains the volume of the elementary cell
$\Delta p \Delta x = \hbar$ or, at least, after integration over momenta in the
case of systems of at equilibrium, the thermal de Broglie wavelength $\Lambda$.
Thus we also have to find how such quantities associated with particles appear
in a FT.

In contrast to the Hubbard-Stratonovitch type descriptions which are finite as
they represent a rigorous mathematical transformation of a finite quantity, we
expect that a field theory, as is often the case, will contain infinities
associated with the short scale spatial discretization of the functional
integral. This will require the introduction of a renormalization procedure.

Finally, to be able to assert that the FT is also an exact representation,
we have to show that there exists a rigorous mapping between the FT and the
standard statistical mechanics of dense systems.
Hereafter we turn our attention to all these questions.

\newpage
\section{Defining the Hamiltonian}\label{sec:definingH}
To build $H[\rho]$ we follow an approach inspired by the methods developped in
QFT where instead of $H[\rho]$ the lagrangian $L[\phi]$ is considered. To find
the latter, we select a functional and check that the mean field approximation
of the theory reproduces a well known result; for instance, the Maxwell
equations in quantum electrodynamics. In this case, $L[\phi]$ can be considered
a good choice for elaborating the complete theory including the fluctuations via
the functional integral.\\

In statistical thermodynamics, it is only for systems without interactions,
ideal systems, that we know an exact and general result and have the explicit
expression of the partition function, $\tilde{\Xi}_{0}$.
We then consider such systems and require that the Hamiltonian
$H_{0}[\rho]$ reproduce the thermodynamic partition function $\tilde{\Xi}_{0}$
in a mean field estimation of (\ref{eq:PartFct0}). 
\textit{However, since $\tilde{\Xi}_{0}$ is a cornerstone in classical statistical
mechanics and contains fundamental physics related to $\Lambda$ and the
indiscernibility of particles, we further require that $H_{0}[\rho]$
reproduce $\tilde{\Xi}_{0}$ exactly, i.e.  also beyond the mean field
approximation.} These fundamental aspects will then be
correctly accounted for in the FT for all systems including those with
interactions. We shall now discuss the Hamiltonian.

Having chosen the field $\rho$ so that its average corresponds to the density
of particles $\tilde{\rho}$, it is evident to fix the chemical potential, $\mu$,
and choose for $\tilde{\Xi}_{0}$ the grand canonical
partition function. In this case we have the exact thermodynamic results
\begin{equation}\label{exact0}
\ln \tilde{\Xi}_{0} = \beta p V = \tilde{\rho}V 
\end{equation}
and
\begin{equation} \label{mu}
\beta \mu = \ln(\tilde{\rho}\Lambda^3)
\end{equation}
where the exact density $\tilde {\rho}$ is uniformly distributed in space. 
A simple Hamiltonian $H_{0}[\rho]$ that reproduces (\ref{mu}) 
in a mean field approximation of (\ref{eq:PartFct0}) is
\begin{eqnarray}\label{eq:hzero}
 \beta H_{0}[\rho] = \int d\mathbf{r} \left\{\rho(\mathbf{r})
  \left[\ln \left(\rho(\mathbf{r})\Lambda^3\right)- 1\right] -
  \beta \mu  \rho(\mathbf{r})\right\}.
\end{eqnarray}
As for the lagrangian $L[\phi]$ in QFT we cannot claim that $H_{0}[\rho]$ 
is unique. However we see that the part 
\begin{equation} 
  F[\rho]= \int d\mathbf{r}\;\rho({\mathbf{r}})
  \left[\ln \left(\rho({\mathbf{r}})\Lambda^3\right)- 1\right] 
\label{fideal}
\end{equation}
of $H_{0}[\rho]$ represents, in the mean field approximation, the free energy of
an ideal system i.e. the kinetic energy and the entropy.
In Section \ref{sec:discussion}, we compare $F[\rho]$ with the DFT (density
functional theory) \cite{DFT1,DFT2,DFT3} where a similar term appears.

The requirement that $H_{0}[\rho]$ gives the exact result entails a more careful
analysis of the functional integral $\Xi_{0}[\rho]$.  In order to calculate
practically (\ref{eq:PartFct0}), we have to introduce in the r.h.s. a lattice
with a spacing $a$. The result will then depend on this parameter.
In the following, our intention is to find the conditions to obtain the
exact thermodynamic result whatever the value of $a$.
The discrete form of $\beta H_{0}[\rho]$ is
\begin{eqnarray} \label{eq:Hdef}
  \beta H_{0}[\rho] &=&
  \sum_{i}^{V/a^3} \rho({\mathbf{r}_i})a^3
  \left[\ln \left(\rho({\mathbf{r}_i})\Lambda^3\right)- 1\right] -
  \beta \mu  \sum_{i}^{V/a^3} \rho(\mathbf{r}_i)a^3
\end{eqnarray}
and the partition function becomes
\begin{eqnarray}
  \Xi_{0}[\rho] &=& \int \prod_{i=1}^{V/a^3}d[\rho(\mathbf{r}_i)a^3] \;
   e^{-\beta H_{0}[\rho]}
\end{eqnarray}
where we have used in the measure the dimensionless quantity
$\rho(\mathbf{r}_i)a^3$.
Due to the local character of $H_{0}[\rho]$, the calculation of $\Xi_{0}$ is
a product of usual integrals, such as 
\begin{eqnarray}\label{eq:local_int_ideal}
  \int d[\rho(\mathbf{r}_i)a^3] \exp\left\{-\rho({\mathbf{r}_i})a^3
  \left[\ln \left(\rho({\mathbf{r}_i})\Lambda^3\right)- 1\right] +
  \beta \mu  \rho(\mathbf{r}_i)a^3\right\}.
\end{eqnarray}
Beyond the saddle point, we have
\begin{eqnarray}
  \ln \Xi_{0}[\rho] = \tilde{\rho} V \left[1+\frac{1}{\tilde{\rho}a^3}
     \psi[\tilde{\rho}a^3]\right]
\label{zerofield}
\end{eqnarray}
where the function $\psi$ given in \ref{app:psi} represents the correction to
the exact thermodynamic result $\tilde{\rho} V$ given in (\ref{exact0}).  
As expected, the correcting term becomes negligible when $\tilde{\rho} a^3$
is large. However, the discretization has introduced a cumbersome
dependence on the lattice spacing $a$, which we would like to dispose of keeping
only physically meaningful terms.  Before discussing this point, we
generalize this result for an external potential.

Having to calculate local
integrals the previous result for the partition function can be easily
extended by changing $\beta\mu$ into $\beta \mu - V^{ext}(\mathbf{r}_i)$ where
the external potential is in temperature reduced units. Instead of
(\ref{zerofield}) we now have 
\begin{eqnarray}\label{eq:XiexpVext}
   \ln \Xi_{0}[\rho,V^{ext}] = 
   \sum_{i}^{V/a^3}{\tilde{\rho}}a^3 e^{-V^{ext}(\mathbf{r}_i)} +
 \sum_{i}^{V/a^3}\psi[{\tilde{\rho}}a^3e^{-V^{ext}(\mathbf{r}_i)}]
\end{eqnarray}
where the last term in the r.h.s. of (\ref{eq:XiexpVext}) is given in
\ref{app:psi}. If \mbox{$V^{ext}(\mathbf{r}_i)\approx 1$}, the
corrective term is still negligeable when ${\tilde{\rho}}a^3$ is large.
Equation (\ref{eq:XiexpVext}) is correct as long as $V^{ext}(\mathbf{r}_i)$ does
not vary rapidly on the distance $a$ which is already large in comparison with
the mean distance between particles ($\approx \tilde{\rho}^{-1/3}$).
This condition is a restriction on the validity of (\ref{eq:XiexpVext}).\\
It is possible to release such constraint and generalize
eq.~(\ref{eq:XiexpVext}) to any external potential by noticing that all physical
terms for this system have a well defined dependence on the lattice spacing,
distinct from the corrections associated with $\psi$. We now take into account
the following quantity
\begin{eqnarray}\label{eq:XiexpVextRen1}
  \ln \Xi_{0}^{R}[\rho,V^{ext}] &=&  \ln \Xi_{0}[V^{ext}] -
   \sum_{i}^{V/a^3} \psi[{\tilde{\rho}}a^3e^{-V^{ext}(\mathbf{r}_i)}]\\
 &=&\sum_{i}^{V/a^3}{\tilde{\rho}}a^3 e^{-V^{ext}(\mathbf{r}_i)}.\label{eq:XiexpVextRen2}
\end{eqnarray}
The renormalized quantity is now equal to its value at 
the saddle point whatever the value of $a$. From an operational 
point of view this result must be understood as follows~:
$\exp\{-\beta H_{0}[\rho]\}$ is a formal expression. It represents an
expansion of the exponential around its saddle point value and in this expansion
terms corresponding to $\psi[\tilde{\rho}a^3e^{-V^{ext}(\mathbf{r}_i)}]$ are
discarded in order to obtain the physical quantities.

In the limit $a\rightarrow 0$, the renormalized grand potential is now a finite
quantity and its value
\begin{eqnarray}
  \ln\Xi_{0}^{R}[\rho,V^{ext}]=\int {\tilde{\rho}}\,e^{-V^{ext}(\mathbf{r})}d\mathbf{r},
\end{eqnarray}
corresponds to the standard statistical mechanics result valid for any
external potential which is independent from $a$.\\
Note that the change of limit due to the presence of a potential is a standard
problem in statistical mechanics as shown in \cite{Hill}. For an ideal system
the so called classical statistical mechanics is obtained in the limit 
$\hbar \to {0}$ whatever the value of $\Lambda$. However, in the presence of an
interaction potential, an extra limit $\Lambda \to {0}$ must be taken in order to
keep all information about the interaction potential.\\  

From the results obtained in this Section we assume that the total Hamiltonian 
will be in the form of
\begin{eqnarray}\label{eq:hun}
 \beta H[\rho] = \beta H_{0}[\rho] +
 \frac{1}{2}\int d\mathbf{r} d\mathbf{r}'\rho({\mathbf{r}})
 \beta v(\mathbf{r}-\mathbf{r'})\rho({\mathbf{r'}})
\end{eqnarray}
where at this stage there is a non-local term due to the presence of the
interaction pair potential $v(\mathbf{r}-\mathbf{r'})$.

Hereafter our main goal is to give an operational meaning to (\ref{eq:hun}) 
as we have already done in the case of $\beta H_{0}[\rho]$. 
To calculate the partition function, we need to expand $\exp\{-\beta H[\rho]\}$.
In QFT the calculation of similar quantities is done by introducing a gaussian
propagator and performing the so-called loop expansion.
Here $H_{0}[\rho]$ is purely local and it is not traditional to give a Feynman
expansion for such a term. We have to find a formal propagator and a loop
expansion associated with $H_{0}[\rho]$ in order to be able to treat the local
and non local part of $H[\rho]$ on the same footing. In the next Section we
shall investigate the properties of $H_{0}[\rho]$ and will show that this
expansion is also fundamental when an interaction potential is present.

\newpage
\section{Feynman expansion of $H_{0}[\rho]$.}
\label{ssec:int_representation}
By using the fact that $H_{0}[\rho]$ is local we have obtained a first expansion
eqs.~(\ref{eq:XiexpVext}) and (\ref{eq:Hdivergentexpansion2J}) of
$\Xi_{0}[V^{ext}]$.
In parallel a second expansion  can be performed with Feynman graphs.
Since both expansions are expressed in terms of ${\tilde{\rho}}a^3$ and
$-V^{ext}$ we may identify term by term the expansion coefficients and
by this method, as we shall see, solve complex problems of combinatory.\\
To have a simple Feynman expansion, we choose a constant field $\rho_{z}$ as a
reference state.  Our choice - obviously not unique - is to take
${\rho}_{z}=e^{\beta\mu}/\Lambda^3$ corresponding to the activity
\cite{hansenbook}\footnote{$\tilde{\rho}$ and $\rho_z$ are identical
in the present case of the system without interactions.}.
This choice combines two interesting points~: $i$) it gives an
expansion in terms of activity which will be useful when comparing in
Section~\ref{sec:equivalence} our results to the Mayer expansion and $ii$) it
leads to a very simple propagator. 

\subsection{Gaussian propagator, perturbative expansion.}
Hereafter we write the field as $\rho(\mathbf{r})=
{\rho}_{z}+\delta\rho(\mathbf{r})$ and we have
\begin{eqnarray}\label{eq:Hexpansion}
  \beta {H}_{0}[\rho,V^{ext}] &=& \beta {H}^{(0)}_0[{\rho}_{z},V^{ext}] +
  \beta {H}^{(2)}_0[\delta\rho]+\beta\delta{H}[\delta \rho,V^{ext}]
\end{eqnarray}
where the first term is constant 
\begin{eqnarray}
  \beta {H}^{(0)}_0[{\rho}_{z},V^{ext}] = 
    - {\rho}_{z} V + \sum_{i}^{V/a^3} V^{ext}(\mathbf{r}_i) {\rho}_{z} a^3.
\end{eqnarray}
The second term is quadratic
\begin{eqnarray}
  \beta {H}^{(2)}_0 = \frac{1}{2{\rho}_{z} a^3} \sum_{i,j}^{V/a^3}
   \delta\rho({\mathbf{r}_i})a^3\;\delta_{ij}\;\delta\rho({\mathbf{r}_j})a^3
\end{eqnarray}
and the Kronecker $\delta_{ij}$ plays formally the role of an \textit{interaction}.
Following the terminology of the QFT we call the quadratic term \textit{propagator}.
The remaining term represents the \textit{coupling} part of the Hamiltonian
given by
\begin{eqnarray}\label{eq:deltaH_expansion}
  \beta \delta {H}[\delta \rho,V^{ext}] &=& \sum_{i}^{V/a^3}
    V^{ext}(\mathbf{r}_i)\delta \rho({\mathbf{r}_i}) a^3
+\sum_{i}^{V/a^3}\sum_{l \ge 3}^{\infty} \frac{(-1)^l(l-2)!}{({\rho}_{z} a^3)^{(l-1)}}
   \left(\frac{1}{l!} [\delta \rho(\mathbf{r}_{i})a^3]^l\right).
\end{eqnarray}
It shows the specificity of the present FT, with an
\textit{infinity} of coupling terms whose coefficients depend  
on a numerical factor and the parameter ${\rho}_{z}a^3$. \\
In order to achieve a diagrammatic expansion we rewrite the partition 
function according to 
\begin{eqnarray}
  \Xi_{0}[V^{ext},J] &=& \int \prod_{i=1}^{V/a^3}d[\rho(\mathbf{r}_i)a^3] \;
   \exp\left\{-\frac{\beta {H}_{0}[\rho,V^{ext}]}{\alpha} +
   \sum_{i}^{V/a^3} J(\mathbf{r}_i) \rho({\mathbf{r}_i}) a^3\right\}
\end{eqnarray}
where ${H}_{0}[\rho,V^{ext}]$ includes the external potential, $J$ is 
a generating field and $\alpha$ is a parameter, formally equal to $1$, 
which is useful to organize the loop expansion
in QFT \cite{ZinnJustin,Amit}.  
We can perform the functional integral, using the gaussian integrals
\cite{Kadanoff,ZinnJustin,Amit,Ma} and express the result formally as
\begin{eqnarray}\label{eq:XiJdef1}
  \Xi_{0}[V^{ext},J] =
   \exp\left\{-\frac{\beta {H}^{(0)}_0[{\rho}_{z},V^{ext}]}{\alpha} \right\}
   \left(\sqrt{2\pi {\rho}_{z}a^3 \alpha}\right)^{V/a^3}\nonumber\\
   \exp\left\{
   -\frac{\beta}{\alpha} \delta {H}\left[\frac{\delta}{\delta J},V^{ext}\right]\right\}
   \exp\left\{\frac{\alpha{\rho}_{z}a^3}{2}\sum_{i,j}^{V/a^3} J(\mathbf{r}_i) \delta_{ij}
    J(\mathbf{r}_j)\right\}.
\end{eqnarray}
The second line introduces the operator obtained by replacing the field
$\delta\rho(\mathbf{r})a^3$ with the derivation operator
$\delta/\delta J(\mathbf{r})$
\begin{eqnarray}\label{eq:couplingH_J1}
  \beta \delta {H} \left[\frac{\delta}{\delta J}, V^{ext}\right] &=&
   \sum_{i}^{V/a^3} V^{ext}(\mathbf{r}_i) \frac{\delta}{\delta J(\mathbf{r}_i)}
    +\sum_{i}^{V/a^3}\sum_{l \ge 3}^{\infty}
      \frac{(-1)^l(l-2)!}{({\rho}_{z} a^3)^{(l-1)}} \left(\frac{1}{l!}
  \left[\frac{\delta}{\delta J(\mathbf{r}_{i})}\right]^l\right).
\end{eqnarray}
This operator is applied to the last term in the r.h.s. of (\ref{eq:XiJdef1})
which is gaussian \cite{Kadanoff,ZinnJustin,Amit,Ma}.
The calculation is performed expanding the operator
$\exp\left\{ -(\beta/\alpha) \delta
{H}\left[\delta/\delta J,V^{ext}\right]\right\}$. 
Taking $J=0$ at the end of the calculation, we select terms with pairs of
derivatives acting on the same quadratic form, which corresponds to the well
known Wick theorem \cite{Kadanoff,ZinnJustin,Amit,Ma}. Note that going
from the density field representation of $\Xi_{0}$ to the generating functional
representation, we invert the kernel of the quadratic form. In the present case,
this is a simple operation which consists in taking the inverse coefficient.

\subsection{Diagrammatic representation}\label{sec:graphrepresentation}
\begin{figure}[ht]
\begin{center}\includegraphics[scale=0.65]{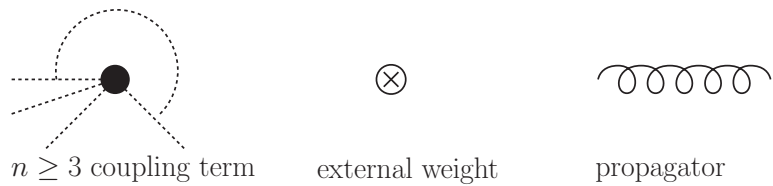}\end{center}
\caption{Diagrammatic elements for the graph representation.}
   \label{fig:diag_elemnt}
\end{figure}

The diagrammatic representation of the theory is organized
\cite{Kadanoff,ZinnJustin,Amit,Ma} around vertices representing
couplings of the field obtained from the expansion of
$\exp\{-\beta \delta {H}/\alpha\}$ and lines joining the vertices
representing the propagator.
The symbols used to draw these elements are shown in figure~\ref{fig:diag_elemnt}.
The propagator is represented by a curly line. The coupling
terms will be denoted by a black circle, where three or more
propagators can be joined, the precise number is understood from the
number $n$ of lines joined to it.
An important feature in knowing explicitly the Hamiltonian functional is that,
as opposed to the phenomenological FT,
we can workout precisely all coefficients for the couplings. Hence, besides the
standard $1/n!$, the coefficient is
$(-1)^{(n-1)}(n-2)!/({\rho}_{z}a^3)^{n-1}$.
The case of the vertex, with only one line attached to it, is drawn by a crossed
circle and is associated with the external weight $-V^{ext}(\mathbf{r})$.
Depending on whether we use the generating functional representation or not,
a coupling term may represent $[\delta\rho(\mathbf{r})a^3]^n$ or
$\left[\delta/\delta J(\mathbf{r})\right]^n$.
Furthermore, otherwise specified, we shall take into account connected graphs
related to the logarithm of the partition function.

Let us now define some topological elements. The \textit{external branches} are the one
body coupling constants together with the only propagator which can be attached
to it.
\textit{Internal lines} are propagators which are not in external branches.

\begin{figure}[ht]
\begin{center}\includegraphics[scale=0.60]{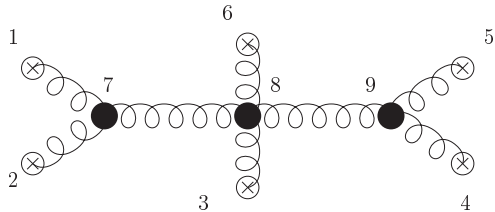}\end{center}
\caption{Example of a diagram.} \label{fig:diag_examp1}
\end{figure}

The graph in figure~\ref{fig:diag_examp1} is an example of a diagram.
The points $7$, $8$, $9$ are vertices, the points $1, ..., 6$ are the external
weights and there are $6$ external branches and two internal lines.

\subsection{Topology.}
\subsubsection{Dimensional analysis.}\label{sec:dimanalysis}
The diagrammatic representation of $\ln \Xi_{0}[\rho]$ leads to an infinity of
graphs that we can classify, as common practice in FT, by using a dimensional
analysis in terms of the parameter $\alpha$ \cite{Kadanoff,ZinnJustin,Amit,Ma}.
This corresponds to the loop expansion.  A diagram with $L$ loops is
dimensionally associated with $\alpha^{L-1}$. For instance, the diagram in
Figure \ref{fig:diag_examp1}, which is a tree diagram ($L=0$), is indeed
proportional to $1/\alpha$.
In the following, our purpose is to show that the dimensional analysis in terms
of the formal parameter $\alpha$ can be associated with a physical
parameter of the system.\\
The standard analysis allows to relating the number of elements in a graph (lines,
vertices) to the number of loops
\cite{Kadanoff,ZinnJustin,Amit,Ma} according to
\begin{eqnarray}\label{eq:loopnumber}
  L - 1 = I + E - m
\end{eqnarray}
where $I$ is the number of internal lines, $E$ the number of external lines, and
$m$ the number of vertices. These include also the one point vertices. The latter
associated with the external potential set the power in $-V^{ext}$.
The r.h.s. of this relation shows that the power in $\alpha$ of the graph
corresponds, in agreement with the expression of the partition function
eq.~(\ref{eq:XiJdef1}), to a factor $1/\alpha$ for each of the $m$ vertices,
$\alpha^{(I+E)}$ for the internal and external lines.
It is tempting to consider the quantity ${\rho}_{z}a^3$ instead of $\alpha$ which
appears in the calculation in a similar way. Here, we also have to account
for the power of this term in each coupling term,
eq.~(\ref{eq:deltaH_expansion}). Let $m_i$ be this power for each of the
$m$ coupling vertices. Considering that each line is attached to two
vertices, we have
\begin{eqnarray}
  \sum_{i=1}^m m_i = 2 I + 2 E.
\end{eqnarray}
Using eq.~(\ref{eq:loopnumber})
\begin{eqnarray}
  \sum_{i=1}^m m_i = 2 (L + m - 1)
\end{eqnarray}
which is also 
\begin{eqnarray}
    -\sum_{i=1}^m (m_i-1) + I + E = - (L - 1)
\end{eqnarray}
On the left hand side we recognize the contribution in powers of ${\rho}_{z}
a^3$ in the graph~: each vertex contributes $1/({\rho}_{z}a^3)^{m_i-1}$, and
there are $I+E$ lines each contributing ${\rho}_{z} a^3$. Thus we obtain a
relation between the overall power of ${\rho}_{z} a^3$ of the graph and the
number of loops.  The role of the parameter for ${\rho}_{z} a^3$ is then
equivalent to that of $1/\alpha$.
In the following, we no longer introduce the factor $\alpha$, as its role is
redundant.
All graphs can be computed exactly and we shall avoid explicit indexing of
the points in the expressions, as finally all points are the same and we shall
only discuss combinatory. The value of a graph is a numerical coefficient,
a power of ${\rho}_{z} a^3$ and of $-V^{ext}(\mathbf{r})$.\\

\subsubsection{Tree graphs.}\label{sec:trees}
First we take the case $L=0$, which corresponds to tree graphs.  From
dimensional analysis, all tree graphs with $n$ external branches are
proportional to ${\rho}_{z} a^3[-V^{ext}(\mathbf{r}_i)]^n$ and the 
value of their sum can be written as
\begin{eqnarray}
 {\rho}_{z}a^3\; c_{n}\; [-V^{ext}(\mathbf{r}_i)]^n
\end{eqnarray}
where $c_{n}$ is a combinatorial coefficient. The value can be obtained by
equating this expression, linear in ${\rho}_{z}a^3$ with the corresponding term
in eq.~(\ref{eq:XiexpVextRen1}) for each point $\mathbf{r}_i$. Thus
\begin{eqnarray}\label{eq:tree_equality1}
  {\rho}_{z}a^3 
   \left(\sum_{n=0}^{\infty} c_n [-V^{ext}(\mathbf{r}_i)]^n\right)
 ={\rho}_{z}a^3 e^{-V^{ext}(\mathbf{r}_i)}.
\end{eqnarray}
Order by order in powers of $V^{ext}$, this equation sets $c_n=1/n!$ for any
$n\ge 3$ and we can generalize the notion of trees to all $n$.
Indeed, one can verify that the cases $n=0$ and $n=1$ relate to the expression of
$\exp\left\{-\beta H[{\rho}_{z}a^3]\right\}$ in eq.~(\ref{eq:Hexpansion}) and
give $c_0=1$ and $c_1=1$ moreover the calculation of the quadratic term in
$V^{ext}$ gives $c_2=1/2$.
Now we know the combinatory for the $n$-tree graphs. 
The result is extremely simple.\\
Of course $c_{n}$ can be also calculated directly by performing the sum of graphs,
such a direct calculation shows that the rather simple and intuitive value of
$c_{n}$ results in fact from the combination of different graphs.\\

\subsubsection{Loop graphs.}\label{sec:loops}
Let us now consider the class of connected diagrams which have at least one loop
($L\ge 1$) and $n$ external branches that we refer to as $n^L$-loop graphs.

For $L>1$, the dimensional analysis states that a given $L$ corresponds to a
power of $1/({\rho}_{z}a^3)$.
We consider graphs with $L>1$ loops and $n$ external branches. For given values
of $L$ and $n$, the dimensional analysis for the sum of all such graphs leads to
the expression
\begin{eqnarray}\label{eq:sum_dL}
   \frac{d_L}{(\rho_{z}a^3)^{(L-1)}} \frac{c'_{n,L}}{n!} 
   [-V^{ext}(\mathbf{r}_i)]^n
\end{eqnarray}
where $d_L$ are the coefficients of the expansion of $\psi$ given
in \ref{app:psi}, and $c'_{n,L}$ is a combinatory coefficient.\\
The case $L=1$ is specific, for $n=0$ we have
\begin{eqnarray}\label{eq:sum_dL2}
  \frac{1}{2}\ln(2\pi{\rho}_{z}a^3)
\end{eqnarray}
and for $n\geq 1$
\begin{eqnarray}\label{eq:sum_dL3}
   \frac{c'_{n,1}}{n!} [-V^{ext}(\mathbf{r}_i)]^n.
\end{eqnarray}

The contributions for any $n$ in eq.~(\ref{eq:sum_dL})~-~(\ref{eq:sum_dL3}) can
be obtained from term by term identification with the function $\psi$ in
eq.~(\ref{eq:Hdivergentexpansion2J}) for all powers of ${\rho}_{z}a^3$
and $-V^{ext}$.
For $L=1$, the comparison gives $c'_{1,1}=1/2$ and for $n\ne 1$ and
$c'_{n,1}=0$. For $L>1$, we must have $c'_{n,L}=(1-L)^n$.
Clearly, the $c'_{n,L}$ can be also calculated by performing the sum of the
corresponding graphs.

\subsection{Ideal system vertex functions.}\label{sec:vertex}
In the following, we define an important object in the diagrammatic expansion.
For $n\ne 2$, we define the $n^T$-vertex functions as the functions obtained from
$n$-tree graphs by erasing the $n$ external branches \footnote{Note that these
vertex functions are not the 1-particle irreducible functions of the field
theory associated with a Legendre transform \cite{Kadanoff,ZinnJustin,Amit,Ma}.}.
The value of the sum of all graphs contributing to a $n^T$-vertex function is
\begin{eqnarray}
 {\rho}_{z}a^3 \frac{1}{n!}
  \left[\frac{1}{({\rho}_{z}a^3)}\frac{\delta}{\delta J(\mathbf{r})}\right]^n
\end{eqnarray}
where $1/({\rho}_{z}a^3)^n$ derives from the fact that we have erased from the
tree graph $n$ external propagators and
$\left[\delta/\delta J(\mathbf{r})\right]^n$ refers to the $n$ points where
this vertex function can be combined to the rest of the graph. The combinatorial
coefficient is that of the corresponding $n$-tree graph. 
The generalisation for the case $n=2$ will be given
later. The general expression is applicable in this case also. The expression of
these tree vertices constitutes an important result of this paper. It states
that despite the variety of graphs contributing to an $n^T$-vertex,
all occurs as if we have a standard coupling of the field at a
given point, with a coefficient which besides the standard $1/n!$, is simply 1.

Starting from graphs with any number $n$ of external branches and loops $L$,
we define the $n^{L}$-vertex functions, obtained similarly to the tree vertex
functions, by removing the $n$ external branches.
For $L=1$ there is a single non zero term for $n=1$
\begin{eqnarray}\label{eq:sum_dL_2b}
  \frac{1}{2}\left[\frac{\delta}{\delta J(\mathbf{r})}\right]
\end{eqnarray}
the other terms for $n \ne 1$ are zero.
And for a given $L>1$ and $n$, the value is given by
\begin{eqnarray}\label{eq:sum_dL_1b}
     (1-L)^n\frac{d_L}{({\rho}_{z}a^3)^{L-1}} \frac{1}{n!}
  \left[\frac{1}{({\rho}_{z}a^3)}\frac{\delta}{\delta J(\mathbf{r})}\right]^n
\end{eqnarray}
where the coefficient is that of the $n^L$-loop graphs. The term in square
brackets is again simply related to the fact that we have removed the $n$
external branches and created the corresponding attaching points.

\subsection{Renormalization.}\label{sec:renormalization}
In the previous analysis, we have associated topological properties of $n$-tree
and $n^L$-loop graphs to given powers of ${\rho}_{z}a^3$ and of $-V^{ext}$.
This has been done in order to relate further this topological analysis with the
analytic expression of the generating functional eq.~(\ref{eq:XiexpVext}).
The sum of tree graphs corresponds to the first term in this equation, whereas
graphs with at least one loop are part of the second term.
As mentionned in Section \ref{sec:definingH}, expression (\ref{eq:XiexpVext})
depends on the lattice spacing whereas the interest of the renormalized
partition function eq.~(\ref{eq:XiexpVextRen1}) is that it has a finite limit
independent of $a$ for vanishing lattice spacing.

Here, in order to free ourselves from the lattice spacing and obtain the
renormalized partition function, we define the following renormalization
procedure which consists in subtracting all graphs with at least one loop.
This is equivalent to susbstracting the term corresponding to the function
$\psi$ in the analytic expression of the partition function, eq.~(\ref{eq:XiexpVext}).
This procedure gives a meaning to the formal expression $\exp\{-\beta H_{0}[\rho]\}$
by giving an operational description in terms of diagrams. Note that after
renormalization, we no longer, strictly speaking, consider $H_0$ and thus
this functional should not be directly compared with other formalisms where it
appears.
From this procedure, we now have a diagrammatic expansion of the renormalized
partition function which corresponds to the exact result for an ideal system and
which can be used for any value of $a$ in particular in the limit of vanishing
$a$, which we discuss in the next Section.

In the following, we shall study the system with interactions and show
that the same graphs as discussed in this Section appear.
We will see that for the reason of locality the renormalization described
here can be applied in this context and that we can obtain a well behaved
theory also for the system with interactions.\\
The present discussion may appear like a cumbersome way of treating the simple
ideal system. However, the crucial point is to understand how the counting
properties for the particles transpose to the FT.
In the following, the main tools introduced in this Section will be
used to analyse the case of a system with interactions, as we are now
able to expand in the same way both local and non local terms in $H[\rho]$ using
the Feynman expansion.

\newpage
\section{Feynman expansion of the full Hamiltonian}\label{sec:InteractingGas}
\begin{figure}[htb]
\begin{center} \includegraphics[scale=0.62]{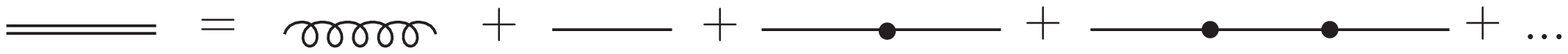}\end{center}
\caption{Diagrammatic representation of the decomposition eq.~(\ref{eq:propdecompos}).}
\label{fig:prop_decompsingle}
\end{figure}

\begin{figure}[htb]
\begin{center} \includegraphics[scale=0.50]{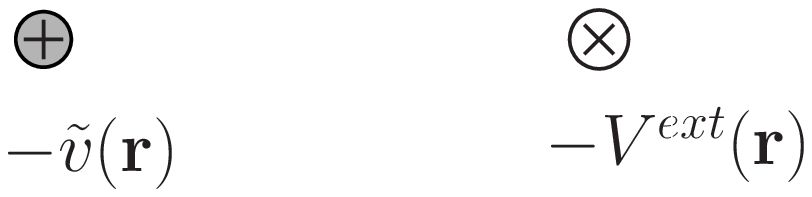}\end{center}
\caption{Diagrammatic notation for the weights at the end of the external branches.}
\label{fig:ext_weight}
\end{figure}

Hereafter we study the generating functional 
\begin{eqnarray}
  \Xi[J] &=& \int \prod_{i=1}^{V/a^3}
   d[\rho(\mathbf{r}_i)a^3] \; \exp\left\{-\beta {H}[\rho]
  + \sum_{i}^{V/a^3} J(\mathbf{r}_i) \rho({\mathbf{r}_i}) a^3\right\}
\end{eqnarray}
in which ${H}[\rho]$ is given in (12). 
Expanding the field around the activity ${\rho}_{z}$, we obtain
\begin{eqnarray}
  \beta{H}[\rho]&=&\beta{H}^{(0)}[{\rho}_{z},V^{ext}]
   +\beta {H}^{(2)}[\delta\rho]+ \beta \delta{H}[\delta\rho,V^{ext}].
\end{eqnarray}
The first contribution is
\begin{eqnarray}
  \beta{H}^{(0)}[{\rho}_{z},V^{ext}]&=&
   -{\rho}_{z}V+\frac{1}{2}{\rho}_{z}V\tilde{v}_{0}
  + {\rho}_{z} \tilde{V}^{ext}_{0}
\end{eqnarray}
where we define
$\tilde{v}_{0}=\beta {\rho}_{z} \sum_{j\ne i}^{V/a^3}v(r_{ij})a^3$
and $\tilde{V}_{0}^{ext}= \sum_{i}^{V/a^3}V^{ext}(r_i)a^3$.
In the following, we assume that we have substracted the self energy and that
the interaction potential cannot be taken at the same point, although to
simplify the notation we do not explicitly indicate it.  The quadratic
Hamiltonian is
\begin{eqnarray}\label{eq:int_propagator}
  \beta {H}^{(2)}[\delta\rho]=\frac{1}{2{\rho}_{z} a^3} \sum_{i,j}^{V/a^3}
   \delta \rho({\mathbf{r}_i})a^3\;[\delta_{ij}+\beta{\rho}_{z}a^3v({r}_{ij})]\;
   \delta \rho({\mathbf{r}_j})a^3
\end{eqnarray}
As noted earlier, the Kronecker $\delta_{ij}$ will be treated as an \textit{interaction}. 
The coupling Hamiltonian is given by
\begin{eqnarray}
  \beta \delta{H}[\delta\rho,V^{ext}]&=&
  \sum_{i}^{V/a^3}\delta \rho(\mathbf{r}_i)a^3 \tilde{v}(\mathbf{r}_i)
  + \sum_{i}^{V/a^3} V^{ext}(\mathbf{r}_i) \delta \rho(\mathbf{r}_i)a^3 \nonumber\\
    &&+\sum_{i}^{V/a^3}\sum_{l \ge 3}^{\infty} \frac{(-1)^l(l-2)!}{({\rho}_{z} a^3)^{(l-1)}}
   \left(\frac{1}{l!} [\delta \rho(\mathbf{r}_{i})a^3]^l\right)
\end{eqnarray}
where $\tilde{v}(\mathbf{r})=\beta \rho_{z}a^3 v(\mathbf{r})$.
We point out that this coupling Hamiltonian is essentially the same as for
$H_0[\rho]$ with the exception of a linear term which includes the interaction
potential. Therefore the topology of the diagrammatic expansion will be
similar to the expansion for $H_{0}[\rho$]. The main modifications are in the
existence of a new contribution to the propagator and to the one
body term. We then have for the generating functional
\begin{eqnarray}\label{eq:XiJintdef}
  \Xi[J] = \exp\left\{-\beta {H}^{(0)}[{\rho}_{z};J]\right\}
   \left(\sqrt{2\pi {\rho}_{z}a^3}\right)^{V/a^3}\\
   \exp\left[
   -\beta \delta {H}\left[\frac{\delta}{\delta J};J\right]\right]
   \exp\left\{-\frac{{\rho}_{z}a^3}{2}\sum_{i,j}^{V/a^3} J(\mathbf{r}_i)
   [\delta_{ij} + \tilde{v}({r}_{ij})]^{-1} J(\mathbf{r}_j)\right\}
  \nonumber
\end{eqnarray}
where, like in Section \ref{ssec:int_representation}, we have substituted
$\delta\rho\,a^3$ by the $\delta/\delta J$ and the notation
$[\ldots]^{-1}$ indicates the inverse. The latter can be expanded according to
\begin{eqnarray}\label{eq:propdecompos}
  [\delta_{ij} + \tilde{v}({r}_{ij})]^{-1} = \delta_{ij} - \tilde{v}({r}_{ij})
  + \sum_k^{V/a^3} \tilde{v}({r}_{ik}) \tilde{v}({r}_{kj}) + \ldots .
\end{eqnarray}
In this expression, the Kronecker $\delta_{ij}$ is its own inverse and the rest
represents a sum of terms of alternate signs constituted with chains of single
potentials.  The diagrammatic representation of this equation is given in
figure~\ref{fig:prop_decompsingle}, where the full propagator appears on the
l.h.s. while on the r.h.s. the curly line is the Kronecker $\delta_{ij}$ and the
lines represent a single interaction potential $-\tilde{v}(r_{ij})$.
We can thus generalize the notion of the tree vertex function of
Section~\ref{sec:vertex} to the two body coupling term associated with a weight
$1/(2{\rho}_{z}a^3)$.
The diagrammatic expansion will be the same as the one given in the previous
section, except that the full double line replaces the curly line and that the
external weight has two contributions shown in figure~\ref{fig:ext_weight}.

\subsection{Topological reduction~: ideal system vertex functions.}\label{sec:topolreduc1}
Hereafter we expand the propagator according to the decomposition shown
in figure~\ref{fig:prop_decompsingle}. The purpose is to apply a topological
resummation of the theory in terms of the vertex functions introduced in
Section~\ref{sec:vertex}.
These vertex functions include at least two attaching points. The case of the
one body coupling term will be detailed separately.

\begin{figure}[ht]
\begin{center} \includegraphics[scale=0.30]{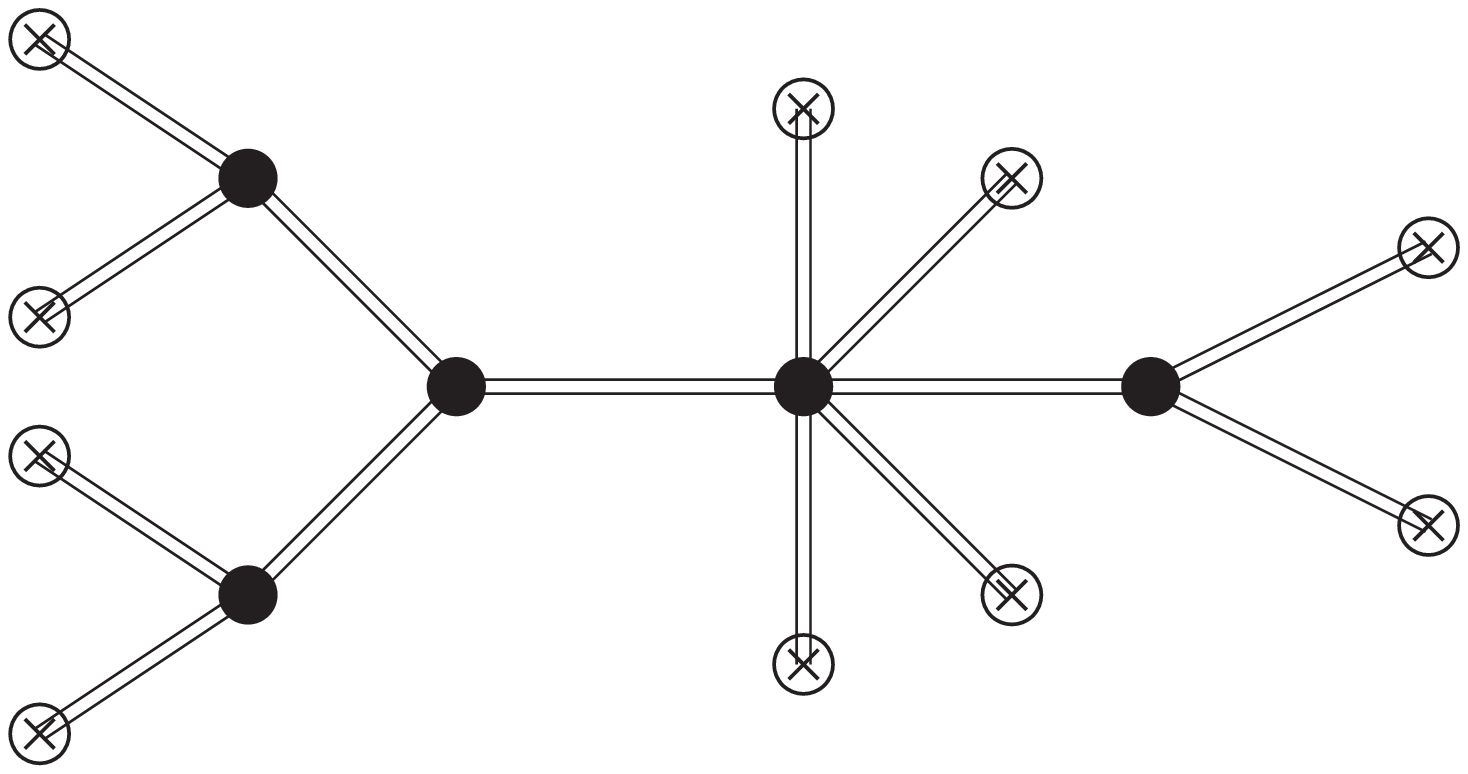}
\hspace{5ex}\Large$\Longrightarrow$\hspace{5ex}
\includegraphics[scale=0.30]{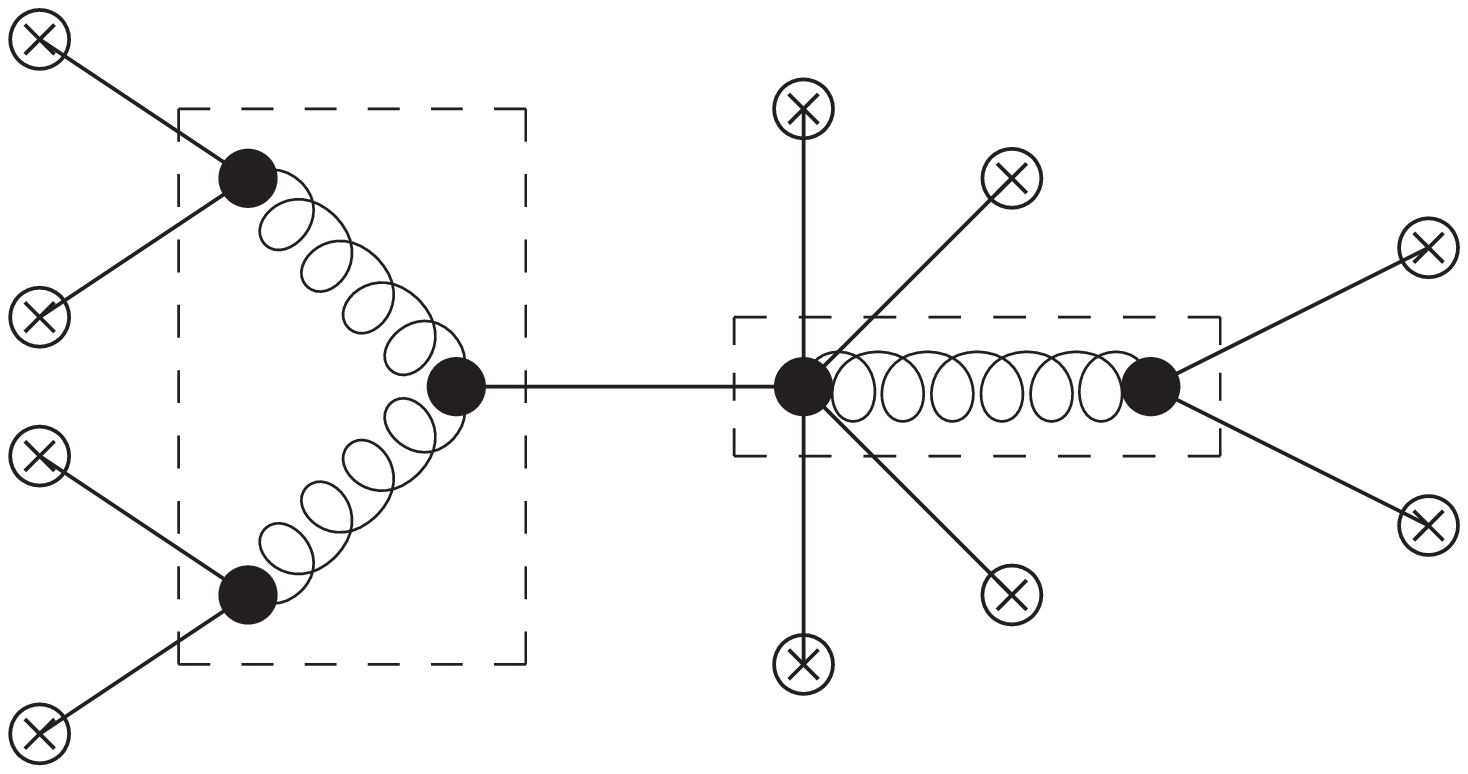}
\end{center}
\caption{Example of a tree diagram with one possible decomposition.}
\label{fig:treev_dbl1b}
\end{figure}

On the graph given in figure~\ref{fig:treev_dbl1b}, we present an example of this
expansion, where the diagram on the right represents a possible decomposition of
the total propagator of the original graph on the left. We have omitted the
labels and arbitrarily chosen one of the external weights $-V^{ext}$.
On the right, for simplicity, we have chosen only the contribution to the chain
of interactions corresponding to a single interaction. These aspects are
irrelevant to the present discussion. Given the local nature of the ideal system
couplings and propagators, it is interesting to isolate in the diagrams the
local parts which are indicated inside the dotted frame on the figure. Their
contribution to the graph is a numerical coefficient as they are independent on
the rest of the graph. 

\begin{figure}[ht]
\begin{center} \includegraphics[scale=0.40]{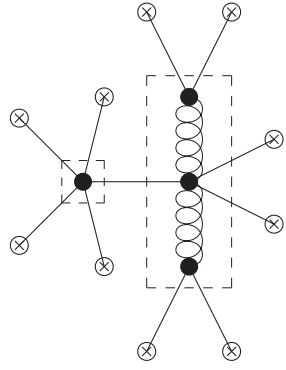}\end{center}
\caption{Diagram which has the same structure for the interactions as the diagram in
figure~\ref{fig:treev_dbl1b}, but a different topology for the ideal propagator.}
\label{fig:treev_mxt1b2}
\end{figure}

\begin{figure}[ht]
\begin{center} \includegraphics[scale=0.50]{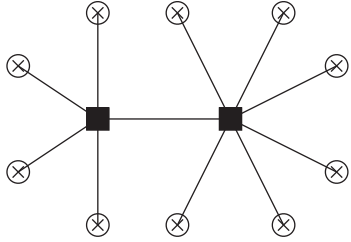}\end{center}
\caption{Representation of the sum of all diagrams with the same structure
for interactions as the diagram in figure~\ref{fig:treev_dbl1b} using the
$5^{T}$-vertex and $7^{T}$-vertex functions. Amongst these we have the diagrams of
figure~\ref{fig:treev_dbl1b} and figure~\ref{fig:treev_mxt1b2}.}
\label{fig:treev_fin1bb}
\end{figure}

\begin{figure}[htb]
\begin{center} \includegraphics[scale=0.50]{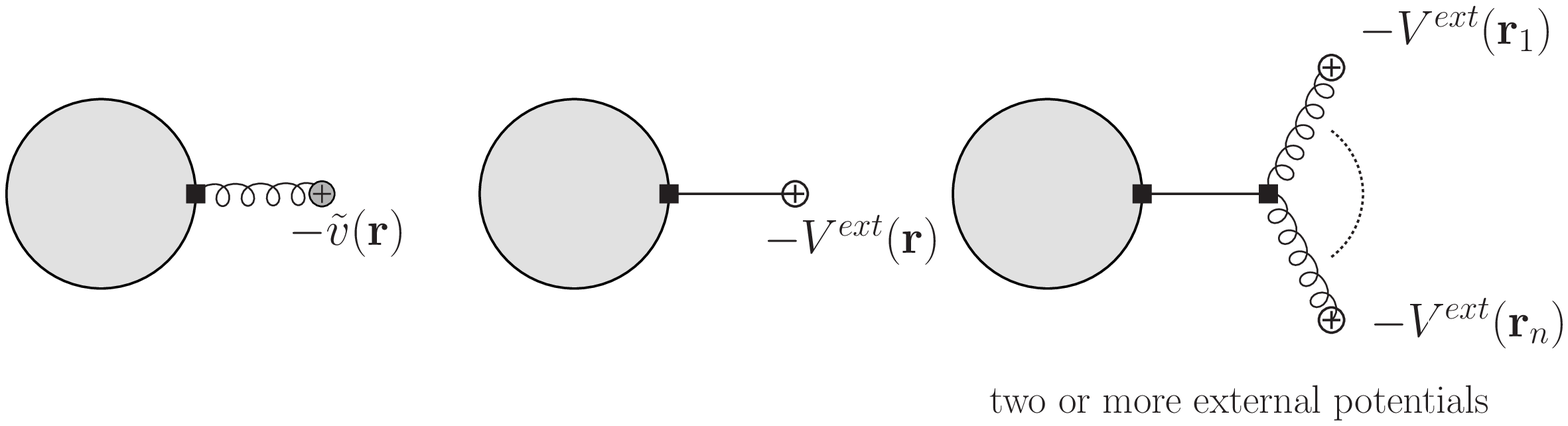} \end{center}
\caption{Resummation of external branches with external potential on the one body
coupling terms. The gray circle may represent any diagram and
the black square a three or higher vertex function.}
\label{fig:v_ext_resum2}
\end{figure}

We then consider graphs with the same backbone structure in terms of the
interaction potentials but with a different local part like, for instance,
in figure~\ref{fig:treev_mxt1b2}.
The sum of all such local diagrams can be performed using the
$n^T$-vertex functions as defined in Section \ref{sec:vertex}.
In the present case, it requires the $5^{T}$-vertex and $7^{T}$-vertex functions
derived from the $5^{T}$-tree and $7^{T}$-tree.
The resummation into vertex functions is equivalent to a topological reduction.
Noting the ideal system vertices introduced in \ref{sec:vertex} by
black squares, the graph is now represented by figure~\ref{fig:treev_fin1bb}.
Clearly, the new graph corresponds to various topologically different graphs
of the original expansion in terms of the total propagator.
The factor associated with each ideal system vertex function is simply
according to Section~\ref{sec:vertex}~: $1/[n!({\rho}_za^3)^{(n-1)}]$.\\
We hereafter detail the special case of the one body coupling which according to
figure~\ref{fig:ext_weight} can have different weights~: the interaction potential
or the external potential.
$i)$~First we discuss the case when the external branch has a weight
$\tilde{v}$.  This weight can be attached on the ideal system propagator or on a
chain of one or more potentials. These two cases are complementary to allow for
any number of interactions in the chain%
\footnote{We recall that from the decomposition in figure~\ref{fig:prop_decompsingle},
there cannot be an ideal and an interaction propagator in series.}.
$ii)$~Then we consider the external branch associated with the external 
potential.
The case when there is a single interaction potential on which we attach the
external potential is specific. It is shown in figure~\ref{fig:v_ext_resum2}
where one may verify that all cases with any number of $-V^{ext}$ are represented.
For the case of a single interaction potential and all other cases, it is
straightforward to find that any coupling term can be decorated with a factor
$e^{-V^{ext}}$.

The topologigal reduction, presented here on a tree graph using the ideal system
tree vertex, can be generalized to graphs which include loops for the ideal
system.
The separation into non local and local parts related with the ideal propagator
can be performed in an identical way. We then also need to introduce the ideal
system loop vertices.
Having performed the topological reduction of all local parts using either tree
vertices or loop vertices, we note that graph with an identical structure
may appear once with a tree vertex and once with a loop vertex.
For a given position of all other vertices, we can combine the tree vertex and a
loop vertex as they are taken at the same point knowing that the rest of the
graph is identical.  
The sum of these two vertices corresponds to the two terms in the ideal system
partition function eq.~(\ref{eq:XiexpVext}). At this stage, we can introduce the
renormalization presented in Section~\ref{sec:renormalization} which corresponds
to substracting the loop vertices. As a result we have vertices which are well
behaved in the vanishing lattice space limit.

As the potential couples distinct points, note that there cannot be a
loop consisting of a single interaction potential. More general loops which may
include chains of interaction potential are not concerned by the topological
reduction associated with the ideal system and remain unchanged.

\subsection{Diagrammatic definition of the partition function.}
\label{sec:diagtotalpartfunction}

The result of this Section is that the logarithm of the partition function is
given by all possible connected graphs made of non labelled ideal system
vertices i.e. $n^T$-vertex functions ($n \ge 2$) and internal lines
corresponding to a single potential. The coefficients of the vertices are those
of the $n^T$-vertices given in Section~\ref{sec:vertex}. The external branches
are either the ideal system or a single potential propagator. At the
end of the external branches we find the labelled weights shown in
figure~\ref{fig:ext_weight}.

\newpage
\section{Mapping between the FT and the Mayer expansion.}
\label{sec:equivalence}

In this Section, we show that our field theory is as thorough as the standard
methods in statistical physics. In order to do this we compare our expansion
with the standard Mayer expansion. To simplify the discussion, we
first consider the case where the external potential is zero.

To elaborate this comparison, we take the standard expansion of the grand
potential in Mayer functions~:
$f(\mathbf{r}) = \exp[-\beta v(\mathbf{r})] -1$ and activity
${\rho}_{z}$ \cite{hansenbook}.
We further expand the exponential in terms of $-\beta v$ and obtain one
and the sum of graphs with $n\ge 1$ lines in parallel representing the
potentials multiplied by a factor $1/n!$.
This corresponds to an expansion in terms of the single potential
introduced in Section~\ref{sec:InteractingGas}, with all possible connected
unlabelled graphs with vertices corresponding to the activity. In the following,
this expansion will be referred to as Mayer expansion. 
The Mayer and Feynman graphs have the same topological elements. They both
include all possible connected graphs made of lines and points. 
To then state the equivalence between the two expansions, we must discuss
the following. 
Firstly, although they have a similar topology, lines and vertices
are associated with different quantities.
In FT, points are $n^{T}$-vertices and one body external weights, whereas in the
statistical mechanics they represent the activity.
Secondly, we need to compare the combinatory coefficients for the two expansions.
Hereafter, we do not discuss the powers of $a$ as in the
final the graph is proportional to $a$ and focus only on ${\rho}_{z}$.

First we discuss the powers in ${\rho}_{z}a^3$.
From the previous section, Feynman diagrams are based on the $n^{T}$-vertex
functions associated with $1/({\rho}_{z}a^3)^{n-1}$.
On these vertices, interaction potential lines are attached which from relation
(\ref{eq:XiJintdef}) each contribute with $({\rho}_{z}a^3)^2$ (one instance
is explicit and one comes from the definition of $\tilde{v}$).
This factor can be distributed on the two vertices to which any line is attached.
By doing this we associate a single power ${\rho}_{z}a^3$ to each vertex and
none to the lines.
The role of the one body vertex has to be treated separately.
In one case, the external weight $\tilde{v}$ is attached to the ideal system
propagator. From eq.~(\ref{eq:int_propagator}), the only factor ${\rho}_{z}a^3$
of the ideal propagator is already distributed to the vertex inside the diagram.
One can verify that there remains one factor ${\rho}_{z}a^3$ associated with
the external weight $\tilde{v}$ and this term corresponds to the activity which
should be at the end of an external line in the Mayer expansion. In the second
case, the external weight is attached to a single interaction potential. This
corresponds to two potentials in series and we can use the two body vertex.
We can verify that here also we have the correct number of factors
${\rho}_{z}a^3$ once they are redistributed on each vertex and that we retrieve
the standard Mayer graph result.
The final statement is that although in the Feynman expansion factors
${\rho}_{z}a^3$ are associated both with lines and vertices, they can formally
be redistributed in order to associate a single instance of this coefficient to
each vertex. This corresponds to the Mayer diagrams expansion where the
activity is associated to the points.

The second aspect is that the $n^{T}$-vertex functions are associated with the
standard $1/n!$ factor. This is exactly the correct combinatory so as to obtain
the non labelled graphs of the Mayer expansion, with identical rules for the
symmetry of the graphs. One only needs to treat separately the case of the
external weight, which includes the interaction potential when it is attached to
the ideal propagator. It corresponds in the Mayer expansion to a single
potential pending from a graph, the topological equivalent of an external branch
in the Feynman expansion. The expected combinatory is found in this case too. 

The sum of all these results shows that the Mayer and the Feynman expansion are
finally identical.
The result can be extended to the system in the presence of an external field,
indeed we have shown that any vertex function can be decorated by a factor
$e^{-V^{ext}}$.
We have seen above that each vertex function can be associated to a factor
${\rho}_{z}a^3$, the multiplication of this factor by the exponential
corresponds, in the liquid state physics, to the generalization of the activity in
the presence of an external field denoted $z^*={\rho}_{z}e^{-V^{ext}}$ in
\cite{hansenbook}.

The foremost result of this paper shows that given the renormalization
introduced in Section \ref{sec:renormalization}, the result for the diagrammatic
expansion is simple and leads to the equivalence of the Feynman and Mayer graph
expansions. We thus fulfil our main objective which is to define a FT capable of
describing the system at a microscopic level introducing a simple and intuitive
Hamiltonian.  
This confirms previous results where we have shown that our formalism reproduces
two exact results which are the virial theorem \cite{virial} and the contact
theorem \cite{ddcjpbjstmolphys2007} which will be discussed in more detail in
the next Section.
Note that in our formalism, there is no reference to Gibbs ensembles.
The difference in number of degrees of freedom associated with the field
description and the lattice spacing $a$ calls for the renormalisation which we
have introduced in order to reproduce the correct combinatorics for particles.

\newpage
\section{Discussion.}\label{sec:discussion}
From standard text books \cite{Hill,FeynmanHibbs}, we know that the so-called
classical statistical mechanics contains two basic properties governed by
quantum physics. Namely, the thermal de Broglie wavelength, $\Lambda$, and the
indiscernibility of particles which originates from $N$ distinct particles a
$N!$ coefficient in the partition function. These elements are not related to
the interaction potential.
In the present paper, we have shown that a simple local functional together with
a renormalization procedure can reproduce these two properties. 
This procedure is not modified when an interaction pair potential is introduced
in the Hamiltonian and consequently we can then demonstrate that the theory is
equivalent to the usual statistical mechanics. 
We have shown that the local functional leads, in perturbation theory, to a
simple combinatory of the fields.
In each monomial term, the $n$ fields are equivalent and their permutation is
associated with the coefficient $1/n!$.  In other words, the local functional
transposes to the FT the indiscernibility of particles.

One characteristic of our FT is that we have been able to introduce a
renormalization procedure through which all the results are finite and
independent from, arbitrary lattice spacing, although there exists an infinity
of coupling constants. 
Due to renormalization, the expression $\exp\{-\beta H[\rho]\}$ is formal and
we must consider that this quantity is defined by its series expansion around
the activity and that some terms in this expansion are cancelled by counter terms;
these are independent of the interaction potential, showing that they have no
physical meaning but are originated only by a mathematical procedure.  

Achieving a microscopically faithful description shows that a simple FT is not
necessarily associated with a coarse graining and can have a level of
description equivalent to that of the standard statistical mechanics, in
contrast to the common conceptions of this type of approach \cite{ReissH}. 
Indeed, the measure we have used does not
require the introduction of any normalization constant in the partition
function, necessary in the case of a coarse grained approach.\\
We can also compare this FT with other microscopically exact field theoretical
descriptions. Considering a field approach without using as a starting point the
standard partition function, we deal with a renormalization that does not exist
for field theories based on the Hubbard Stratonovich transform.  On the other
hand, our field is extremely simple and has an obvious physical meaning. This
contrasts with the Hubbard Stratonovich type approaches, where we have to work
in a complex plane with an auxiliary field for which it is rather difficult to
introduce appropriate physical approximations.    

We also emphasize that FT is distinct from the DFT.
Both approaches are based on the existence of a functional of the density.
However, in the two formalisms, the correlations are treated in
different ways \cite{ReissH,Evans}.  
In the DFT, the form of the functional includes all correlations and
fluctuations and we know that this functional exists but we ignore its exact
form.  Minimizing this functional yields the equilibrium density distribution.
In contrast, in FT the functional is known and simple.
The core of the FT formalism is to gradually account for the fluctuations when
calculating quantities for the system in a perturbative expansion. One part of 
$H[\rho]$ is $F[\rho]$ which is formally like the free energy of the ideal system. 
A similar term $F_{DFT}[<\rho>]$ is introduced in DFT, however it is important to point
out the differences between $F[\rho]$ and $F_{DFT}[<\rho>]$. $F[\rho]$ 
is a functional of a field i.e. a fluctuating quantity where as, at the minimum,
$F_{DFT}[<\rho>]$ is a function of the mean value of the fluid density.
Moreover, we have mentionned earlier that $\exp\{-\beta H[\rho]\}$ is essentially
a formal expression.
This illustrates one specificity of the FT~: the fluctuations of the ideal term 
which basically represent the entropy must be considered on the same footing 
as the fluctuations related to the interaction pair potential. \\
In this respect, although our FT is equivalent to standard statistical
mechanics, the two approaches focus on different aspects of the correlations.
This is the case when comparing standard approaches, but the discussion will be
extended below for the case of our FT.
We are convinced that having at disposal distinct formulations for a given
quantity is indeed useful, possibly for acquiring a broader understanding.

\subsection{Examples}
Hereafter we illustrate on three examples how FT leads to a new point of
view on traditional quantities.\\

In liquid state theory there are three classical expressions of the chemical
potential. One of them corresponds to \cite{RowlinsonWidom}
\begin{eqnarray}
\ln(\tilde{\rho}(\mathbf{r})\Lambda^3) + \ln <\exp(\beta u(\mathbf{r}))>
  + V^{ext}(\mathbf{r})&=& \beta \mu.
\end{eqnarray}
A second traditional expression is given by 
\cite{RowlinsonWidom,MoritaHiroike,StillingerBuff}
\begin{eqnarray}\label{eq:one_ms}
  \ln(\tilde{\rho}(\mathbf{r}) \Lambda^3)
   -c^{(1)}(T,[\rho];\mathbf{r})+V^{ext}(\mathbf{r})=\beta\mu
\end{eqnarray}
where $c^{(1)}(T,[\rho];i)$ is the single-particle direct correlation function
\cite{hansenbook,MoritaHiroike,StillingerBuff}.
Finally, we also have a relation based on a charging process of the interaction
potential \cite{Hill}
\begin{eqnarray}\label{eq:one_charging2}
  \ln (\tilde{\rho}(\mathbf{r}) \Lambda^3) +
   \tilde{\rho}(\mathbf{r})\int_0^1d\xi \int d\mathbf{r}'
   \beta v(|\mathbf{r}-\mathbf{r}'|) g^{(2)}(|\mathbf{r}-\mathbf{r}'|;\xi)
    + V^{ext}(\mathbf{r})
   = \beta \mu
\end{eqnarray}
where $g^{(2)}(r_{ij};\xi)$ is the pair distribution function \cite{hansenbook}
as a function of the charging parameter~$\xi$.
We note that all these expressions emphasize properties related to the
potential, whether calculating the correlations of a quantity involving the
interaction, or calculating the single-particle direct correlation function
or alternatively considering a charging process of the interaction.

The field theoretical description leads to a new expression which can be
obtained by writing that the field is a dummy variable in the functional
integral. This leads to the so called Dyson relations
\cite{FeynmanHibbs,slovenia} and we obtain
\begin{eqnarray}
  <\ln(\rho(i)\Lambda^3)> +
    \sum_{j; j\neq i}^{V/a^3} \beta v(i,j) <\rho(j)> a^3
   + V^{ext}(i)&=&\beta \mu.\label{eq:dyson1b}
\end{eqnarray}
Here the term related to the interactions is rather simple, it expresses the
mean potential at a given point without taking into account the correlations. 
All correlations and fluctuations appear in the calculation of the average of
the logarithm of the density field. This contrasts with a simple term like the
logarithm of the average density, which appears in standard statistical
mechanical expressions or in the DFT.
As a consequence, differences in the description and a different organisation of
the perturbation expansion in the FT, suggest that one should be able to
elaborate new approximations.\\

Let us consider now the so-called contact theorem which establishes an exact
relation between the pressure $p$ existing in a bulk phase and the value of the
density profile $\rho(0)$ at the wall enclosing the bulk material. 
This corresponds to
\begin{eqnarray}
  \beta p = \rho(0).
\end{eqnarray}
In so far as this relation is concerned, discussing the derivation of this
theorem is an opportunity to emphasize the conceptual differences between the
various approaches.
We mention the kinetic theory of gases, in which this relation is the
consequence of the mechanical equilibrium at the interface.
In the case of DFT, the derivation is straightforward
as we only need to write a displacement of the external
potential~: the interface, in two different ways.
Another derivation \cite{JRHenderson} is obtained by integrating the BGY
equations. In this case, a \textit{subtle} integration of the correlations through the
interface leads to the relation.
Within our field theoretical framework, the key element is the local functional
which is essential at different levels. It is crucial to obtain the
density contact value present in the contact theorem
\cite{ddcjpbjstmolphys2007}, but it is also necessary to cancel supplementary
terms which appear in the demonstration.  In this respect, specific relations of
the field theory are also required, namely the Dyson type relations
\cite{slovenia}.\\

Now, in a third example, we illustrate one of the main aspects of the FT, i.e. 
the existence of an intricate coupling between counting (entropy) and
interaction.  Let us study the interfacial properties of ionic fluids. 
From the point of view of the interactions, we know that the important quantity
is the charge, the difference of densities of each species.
However, this system can also be viewed as a peculiar mixture which has
a specific condition due to electroneutrality.
From this point of view, we have two terms
in the ideal functional describing the indiscernibility for each ion.
Thus the natural fields are the densities describing each ion.
In \cite{ddcjstjpbMolPhys2003a,ddcjstjpbElectrochimActa}, we show, in the
specific instance where the natural fields for the ideal and for the interaction
term are distinct, that the perturbation theory leads to a coupling of the charge and
of the total density field due to the local ideal functional.
This has direct consequences.\\
For the simple neutral interface, we show that there exists a depletion for the
quadratic fluctuations of the charge.
Then, the entropic coupling between the charge field and the total density field
predicts a non trivial profile on the total density
\cite{ddcjstjpbMolPhys2003a,ddcjstjpbElectrochimActa}.
We can verify that the contact value of this total density profile satisfies the
exact condition of the contact theorem, for the pressure calculated at the same
level of approximation.
We have used this phenomenon to analyse the anomalous behaviour of the differential
capacitance as a function of the temperature \cite{anomalous1,anomalous2} which
has been thoroughly discussed recently in experiments
\cite{Tosi1,Tosi2}, numerical simulation \cite{BodaChan} and
theoretical approaches \cite{MierYTeran,BodaHolovko,Sokolowski,Outhwaite1}.
The interest of our analysis is that it provides a simple interpretation and
understanding for this phenomenon, associating the decrease in the capacitance
with the depletion of the ionic density at the interface at low temperature and
providing the physical origin of this depletion.

Moreover, the more detailed account of these entropic effects is fundamental in
the case of asymmetric, in valence, electrolytes.
In \cite{anomalous2}, we have tested our FT by comparing with the results of
numerical simulations \cite{bodaasym} and shown that the theory accounts for all
main qualitative properties of the phenomenon, in comparison with other approaches
\cite{Outhwaite2} which although currently more quantitative fail to take certain
features into consideration.

\newpage
\section{Conclusion}\label{sec:conclusion}
In this paper, we present a field theory describing classical fluids at
equilibrium at the same level as the standard statistical mechanics.
We introduce a real physical field and construct the Hamiltonian
in the spirit of the QFT.
This functional includes interactions and a local functional representing
the ideal system.
The latter characterises our approach and has been thoroughly discussed.
In particular, we show that it provides, for the FT, essential ingredients
in relation to quantum mechanics.
The equivalence of our theory with standard statistical mechanics is shown
by establishing that the Feynman expansion of the FT is equivalent to the
standard Mayer expansion.
The approach is original in that it is not a simple mapping of the standard
partition function like other field theories. Consequently, it requires
a renormalization which we describe.
Its basic interest is that the theory remains simple and intuitive
like phenomenological field theories.

Establishing a field theory which is both a simple and exact representation of
the statistical mechanics has many advantages. We present possible applications.\\
Some are related to the FT formalism. We can, for instance, use powerful
tools such as discussions in terms of symmetries of the system, of the fields
\cite{desorption}.
Also, the fact of having a field variable at the microscopic level should allow
for natural bridging with the mesoscopic intuitive approaches which also adopt
the field theory description.
An example can be found in \cite{IonicFT2} where a mesoscopic Hamiltonian
is presented.\\
Another aspect is that this formalism treats fluctuations in a different way.
This type of approach would help elaborating small systems, where
fluctuations can have the same magnitude as the quantities characterising the
system \cite{ReissH}.\\
Finally, we have also shown that there is an emphasis on correlations
associated with entropic effects.  Such emphasis should shed new light on the
description of ionic systems, or mixtures.
For instance, we believe that the emphasis on the correlations between charge
and total density could add to the understanding of criticality in ionic
systems.  For such systems, the potential couples the charge, whereas criticality
characterizes a phenomenon on the total density. 
Another system of interest in the field of the double layer is the study of
asymmetric in charge electrolytes, which exhibit polarization phenomena even in
the vicinity of neutral interfaces.
As opposed to asymmetric in size ions, this phenomenon
is not intuitive. The difference of density of anions and
cations for these asymmetric systems seems to be the origin of such phenomena as
a consequence again of entropic effects \cite{zasymnew}.\\

\section*{Acknowledgements}
The authors would like to thank Dr. J. Stafiej for stimulating discussions and
comments.

\newpage
\appendix
\section{Beyond the ideal system saddle point}\label{app:psi}
Beyond the saddle point, we can compute the integral
eq.~(\ref{eq:local_int_ideal}) taking into account the fluctuations of the
field, on each lattice site we expand the density field as
$\rho=\tilde{\rho} +\delta\rho$, in this case the logarithm of the partition
function is
\begin{eqnarray}
  \ln \Xi_{0} = \tilde{\rho} V + \frac{V}{a^3}\ln\left[\int_{-\infty}^{\infty}
  dt \, e^{-t^2/2}
  \exp\left(\sum_{n=3}^{\infty} \frac{(-1)^{n+1}t^n}
    {n(n-1)(\tilde{\rho} a^3)^{n/2-1}}\right) \right]
\end{eqnarray}
where $t=(\rho-\tilde{\rho})a^3/\sqrt{\tilde{\rho}a^3}$.
Expanding the last exponent, which makes sense in the limit of large
$\tilde{\rho}a^3$, we find we have to calculate Gaussian integrals:
$\int_{-\infty}^{\infty} t^{2n} e^{-t^2/2} dt = \sqrt{2\pi} (2 n -1)!!$.
The result can be written
\begin{eqnarray}\label{eq:Hdivergentexpansion}
  \ln \Xi_{0} = \tilde{\rho} V + \frac{V}{a^3}\psi[\tilde{\rho}a^3]
   = \tilde{\rho} V \left[1+\frac{1}{\tilde{\rho}a^3}
     \psi[\tilde{\rho}a^3]\right]
\end{eqnarray}
with
\begin{eqnarray}\label{eq:Hdivergentexpansion2}
 \psi[\tilde{\rho}a^3]&=& \frac{1}{2}\ln(2\pi\tilde{\rho}a^3)
  + \sum_{L=2}^\infty \frac{d_L}{(\tilde{\rho}a^3)^{(L-1)}}\\
  &=&\frac{1}{2}\ln(2\pi\tilde{\rho}a^3)
 - \frac{1}{24} \frac{1}{(\tilde{\rho}a^3)}
 - \frac{1}{48} \frac{1}{(\tilde{\rho}a^3)^2}
 - \frac{161}{5760} \frac{1}{(\tilde{\rho}a^3)^3} + \ldots
\end{eqnarray}
with the exclusion of the first term, $\psi$ is a power series of
$1/\tilde{\rho}a^3$ which is asymptotically convergent for large $\tilde{\rho}a^3$,
for which the first values of the coefficients $d_L$ are given on the second line.\\
The expression in the presence of an external potential $V^{ext}(\mathbf{r}_i)$ is
\begin{eqnarray}\label{eq:Hdivergentexpansion2J}
 \psi[\tilde{\rho}e^{-V^{ext}(\mathbf{r}_i)}a^3]&=& \frac{1}{2}\ln(2\pi\tilde{\rho}a^3)
  -\frac{1}{2}V^{ext}(\mathbf{r}_i)
  +\sum_{L=2}^\infty \frac{d_L \,e^{-(1-L)V^{ext}(\mathbf{r}_i)}}{(\tilde{\rho}a^3)^{(L-1)}}
\end{eqnarray}

\newpage
\section*{References}
\bibliographystyle{iopart-num}

\end{document}